\documentclass[journal]{IEEEtran}

\usepackage{cite}

\usepackage[dvips]{graphicx}
 \graphicspath{{/Figure/}}
\DeclareGraphicsExtensions{.eps}

\usepackage{array}
\usepackage{multirow}
\usepackage{color}
\usepackage{longtable,balance}
\usepackage{graphicx}
\usepackage{supertabular}

\usepackage{epic}
\usepackage{tabularx}
\usepackage{subfigure}
\usepackage{booktabs}
\usepackage{amsmath}
\usepackage{amssymb}
\usepackage{amsopn}
\usepackage{enumerate}
\usepackage{eso-pic}

\begin{document}

\title{From the Internet of Information to the Internet of Intelligence}

\author{F.~Richard~Yu,~\IEEEmembership{Fellow,~IEEE}\\
\IEEEauthorblockA{School of Information Technology, Carleton University, Ottawa, ON, Canada}\\
}

\maketitle

\begin{abstract}
In the era of the Internet of information, we have gone through layering, cross-layer, and cross-system design paradigms. Recently, the ``curse of modeling" and ``curse of dimensionality" of the cross-system design paradigm have resulted in the popularity of using artificial intelligence (AI) to optimize the Internet of information. However, many significant research challenges remain to be addressed for the AI approach, including the lack of high-quality training data due to privacy and resources constraints in this data-driven approach. To address these challenges, we need to take a look at humans' cooperation in a larger time scale. To facilitate cooperation in modern history, we have built three major technologies: ``grid of transportation", ``grid of energy", and ``the Internet of information". In this paper, we argue that the next cooperation paradigm could be the ``Internet of intelligence (Intelligence-Net)", where intelligence can be easily obtained like energy and information, enabled by the recent advances in blockchain technology. We present some recent advances in these areas, and discuss some open issues and challenges that need to be addressed in the future.

\end{abstract}


\IEEEpeerreviewmaketitle

\section{Introduction} \label{sec:introduction}

The Internet has become one of the major foundations for our socio-economic systems by enabling information exchange among people and machines. In the era of the Internet of information, we have gone through layering, cross-layer, and cross-system design approaches. \emph{Layering} in the original design of the Internet is one of the key reasons behind the success of the Internet \cite{CLC07}. In the layering design paradigm, the complex task of end-to-end networked interactions is broken into disjoint parts to simplify the design. Each layer only needs to interact with its adjacent protocol layers.

Although layering can simplify network design and operation, there are a number of issues with the layered architecture \cite{SM05}. For example, layering introduces inefficiencies and/or redundancy (the same function is performed at multiple layers). In addition, it is difficult to share the information about operation at one layer with the other layers. Consequently, layering can lead to poor performance, especially for applications with hard quality of service (QoS) constraints \cite{AEM16}.

To address the issues with the layered architecture, \emph{cross-layer} design became a popular approach at the beginning of the 21st century \cite{SM05, FGA08}. Cross-layer design allows communication to take place between nonadjacent layers in the system. By exchanging information between nonadjacent layers, each layer can make efficient adaptation according to other layers' status. Consequently, substantial gains in throughput, efficiency, and QoS can be achieved with cross-layer design.

While cross-layer design can improve the networking performance, focusing on networking alone is difficult to meet the demands raised by new applications, such as augmented reality (AR) and virtual reality (VR), which require not only high data rate, but also high storage and computation capabilities \cite{CYH16}. Recently, \emph{cross-system} design has attracted great interests from both industry and academia \cite{YHL18}, where networking, caching and computing are jointly considered. The complexity of cross-system design is very high, which makes it difficult to use traditional methods to optimize the whole system due to the ``curse of modeling" and ``curse of dimensionality" \cite{WHY18}. Recent advances in artificial intelligence (AI) have been extensively used in the optimization of cross-system design \cite{YH18}. 

\begin{figure}[tp]
  \centering
	\vspace{-11 cm}
  \includegraphics[width=0.58 \textwidth]{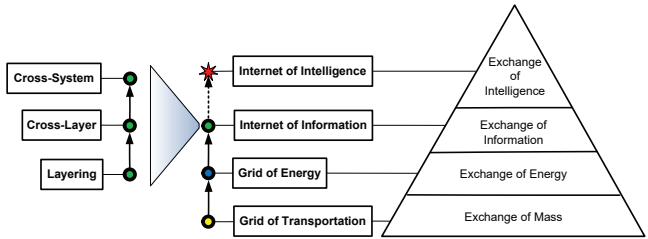}
  \caption{To facilitate cooperation in modern history, we have built three major technologies: ``grid of transportation", ``grid of energy", and ``the Internet of information". Layering, cross-layer, and cross-system designs are  parts of design approaches in the era of the Internet of information. We envision that the next cooperation paradigm could be the ``Internet of intelligence (Intelligence-Net)."}
  \label{fig:History}
\end{figure}

Nevertheless, many significant research challenges remain to be addressed for the AI approach to cross-system design. For example, training data has significant impacts on the AI approach, but high-quality training data may not be available to the system designer due to privacy and resources constraints \cite{Jordan15}. To address these challenges, we need to take a look at humans' cooperation in a larger time scale. To facilitate cooperation in modern history, we have built three major technologies: ``grid of transportation", ``grid of energy", and ``the Internet of information". Figure \ref{fig:History} shows this process, where layering, cross-layer, and cross-system designs are just parts of design approaches in the era of the Internet of information.

In this paper, we envision that the next cooperation paradigm could be the ``Internet of intelligence (Intelligence-Net)", where intelligence can be easily obtained like energy and information, enabled by the recent advances in blockchain technology. We believe that the Intelligence-Net can have significant impacts on the socio-economic systems. We present some recent advances in these areas, and discuss some open issues and challenges that need to be addressed in the future.

%For example, the application layer can inform the lower layers about the application data characteristics and requirements, and lower layers can inform the application player about the nextwork/link/channel conditions. 

The rest of the article is organized as follows. First, we review the design approaches in the Internet of information in Section II. Section III discusses the enabling technologies for humans' cooperation. The Intelligence-Net is presented in Section IV. Section V discusses some open issues and research challenges. Finally, we conclude this article in Section VI.

\section{Design Approaches in the Internet of Information}\label{sec:issue}

In the era of the Internet of information, we have gone through layering, cross-layer, and cross-system design approaches. We describe these approaches in this section.

\subsection{Layering}

It is widely recognized that the general principle of layering is one of the key reasons for the great success of the Internet \cite{CLC07}. Layered architectures form the fundamental structure of the Internet. In layered architectures, each module, called layer, is responsible for specific functions by observing a set of its own parameters and the parameters from other layers, and controlling a set of decision variables. Each layer provides a service to the layer above and hides the complexity of the layer below. 

Layering has several advantages, including simplification, modularity, abstract functionality, and reuse.
\begin{itemize}
\item Simplification: Layering can simplify the design by breaking the complex tasks of end-to-end networking into disjoint parts.
\item Modularity: Each layer is easier to design, develop, optimize, manage, and maintain.
\item Abstract functionality: Each layer can be changed without affecting other layers.
\item Reuse: The functionality and service provided by each layer can be reused.
\end{itemize}

Nevertheless, layering has several disadvantages, including suboptimal, information hiding, and performance.
\begin{itemize}
\item Suboptimal: Layering can introduce inefficiencies and/or redundancy.
\item Information hiding: It is difficult for the information at one layer to be used by other layers.
\item Performance: Layering can lead to poor performance, especially for the applications with hard QoS constraints.
\end{itemize}

\subsection{Cross-Layer Design}
The layered architecture has been under close scrutiny from the research community, especially from the wireless networking research community \cite{SM05,FGA08}. A large number of cross-layer design proposals have been presented in the literature. Cross-layer design actively exploits the dependence between different layers to improve the overall system performance. This is different from layering, where different layers are designed independently, and direct communication between nonadjacent layers is forbidden.

The well-known case of transmission control protocol (TCP) over wireless networks is one of the most commonly cited cases of cross-layer design. In the layered architecture, a TCP sender mistakes a packet error on a wireless link to be an indicator of network congestion, which can result in deteriorated TCP performance. By contrast, new interfaces from lower layers to the transport layer can be created in cross-layer design to enable explicit notifications, by which the performance of TCP can be improved significantly \cite{SM05}. Many other examples of cross-design can be found in \cite{SM05,FGA08,FXD14,AEM16}.

\begin{figure}[tp]
  \centering
	\vspace{-0 cm}
  \includegraphics[width=0.37 \textwidth]{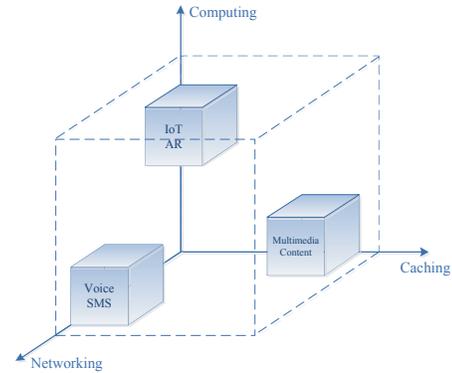}
  \caption{Networking, caching and computing cross-system design for different applications. IoT: the Internet of Things. AR: Augmented Reality. SMS: Short Message Service.}
  \label{fig:Cube}
\end{figure}

\subsection{Cross-System Design}

With the new emerging applications, such as AR and VR, conventional networks solely focusing on networking are no longer capable of meeting the demand raised by such applications not only on high data rate, but also on high caching and computing capabilities. On the other hand, recent advances in information and communication technologies (ICT) have fueled a plethora of innovations in various areas, including networking, caching and computing. 

The integration of these systems becomes a natural trend \cite{YHL18}. By incorporating caching functionality into the network, the system can provide native support for highly scalable and efficient content retrieval. In addition, the integration of computing functionality endows the network with the powerful capability of data processing, hence enabling the execution of computationally intensive applications in the network. By offloading mobile devices' computation tasks to edge/cloud nodes, the local resources of mobile devices especially battery life can be conserved, and the task execution time can be reduced, thus enriching the quality of experience of users. In addition, networking, caching and computing functionalities can complement each other by cross-system design. For example, some of the computation results can be cached for future use, and backhaul workload can be alleviated. Figure \ref{fig:Cube} shows networking, caching and computing cross-system design for different applications.

\subsection{Artificial Intelligence (AI) Approach}

The complexity of cross-system design is very high when we jointly consider the dynamics of different systems together. It is difficult to use traditional optimization methods to the the optimization problem in cross-system design due to the ``curse of modeling" and ``curse of dimensionality". The ``curse of modeling" refers to the difficulty of accurately knowing all the state parameters and dynamics when the modeled system is complex. The ``curse of dimensionality" refers to the difficulty of finding the optimal policy when the state parameters, actions and dynamics increase due to the high requirements of computational time and resources.

Reinforcement learning (RL) \cite{SB18} is a widely used method that can mitigate both the ``the curse of modeling" and the ``curse of dimensionality" in ways that allow us to approximate optimal policies for sequential, state-dependent decision problems when the system dynamics are not explicitly known  and/or when the  problem is highly complex. Nevertheless, RL is unstable or even divergent when action value function is approximated with a nonlinear function like neural networks.

Recent advances in deep learning \cite{GBC16} result in deep RL (DRL), where these two kinds of learning combine nicely. Basically, RL needs methods for approximating functions from data to implement all of its components, including value functions, policies, world models, state updaters, and deep learning is the latest and most successful of recently developed function approximator. DRL has become one of the most important methods for AI, and has been successfully used to solve the optimization problems in cross-system design \cite{YH18}.

Many significant research challenges remain to be addressed for the AI approach to cross-system design. For example, training data has significant impacts on the AI approach, but high-quality training data may not be available to the system designer due to privacy and resources constraints\cite{Jordan15}. 

\section{Enabling Technologies for Humans' Cooperation} \label{sec:enabling}
To address the  challenges described above, we need to take a look at humans' cooperation in a larger time scale. Cooperation is at the core of humans' history. It is believed that humans came to dominate the world because we are the only animal that can cooperate flexibly in a large number \cite{Harari14}. In modern history, to facilitate cooperation, we have built three major technologies: ``grid of transportation", ``grid of energy", and ``the Internet of information", which are described in the following. Figure \ref{fig:History} shows this process.

\subsection{Grid of Transportation}
Transportation technologies have played a vital role in our cooperation, including trade, war, and social activities. Before any other form of transportation, humans traveled on foot. Then, humans learned to use animals and boats for transportation. In the 17th and 18th centuries, many new transportation technologies were invented such as bicycles, trains, motor cars, trucks, airplanes, and trams. Aircraft, space ships and cars are some of the defining transportation technologies of the twentieth century.

Essentially, the primary purpose of transportation is to move mass from one location to another location to reduce the disparity of resources. For example, one country with no oil can import oil from another country using transportation.

\subsection{Grid of Energy}
Another major innovation is the energy grid, which is fundamental to modern life. With the energy grid, we can easily get energy to power our phones and computers, cool our homes, and get light at night, by simply plugging into the electric grid. The primary purpose of the energy grid is to move energy from one location to another location to reduce the disparity of energy. Interestingly, there is a well-known relationship between energy and mass found by Einstein as follows.
\begin{equation}
E = m c^2,
\end{equation}
where $E$ is the energy, $m$ is the mass, and $c$ is the speed of light.

\subsection{The Internet of Information}
Following the energy grid, the invention of the Internet of information has enabled humans' cooperation to the new level. The Internet of information has become one of the major foundations for our socio-economic systems by enabling information exchange. The primary purpose of the Internet of information is to move information from one location to another location to reduce the disparity of information. 

The connection between information and energy can be tracked back to the Maxwell's `demon' \cite{MNV09}. The demon is able to decrease the entropy of the system by converting information (about the position and velocity of each particle) into energy. Landauer's principle  demonstrates that the number of bits of information, $I$, that can be irreversibly erased or merged by an ideally efficient memory change or logic operation require a minimum energy, $E$.

\begin{equation}
I = \frac{E}{T k_B ln(2)},
\end{equation}
where $I$  is the amount of information in bits, $E$ is in physical Joules, $T$ is the temperature, and $k_B$ is the conversion factor from energy in Kelvins to Joules \cite{BAP12}.

From the description above, we can see that each major innovation enables us to move ``something" to reduce the disparity of ``something", by which we can share ``something" to facilitate our cooperation. Here, ``something" is mass, energy, and information in these enabling technologies, respectively. In addition, we can observe that every novel technology is created from existing ones, and every technology stands upon a pyramid of others.

\section{The Internet of Intelligence (Intelligence-Net)}
We envision that the next cooperation paradigm could be the ``Internet of intelligence (Intelligence-Net)", where intelligence can be easily obtained like energy and information, enabled by the recent advances in blockchain technology. In this section, we first present the limitations of the current Internet of information. Then, we present recent advances in blockchain technology, which can be used to enable the Intelligence-Net.

\subsection{Limitations of the Existing Internet of Information and AI}
The traditional Internet was originally designed to handle the exchange of information, e.g., using emails and websites. It was  not designed to handle the exchange of valuable things, including money and intelligence. Although online e-commerce is very popular nowadays, transferring money online is not actually moving the money directly. Instead, intermediary is used, such as a bank, a credit card company, Western Union or PayPal.

Current AI algorithms generally involve a large volume of data. With more data to analyze, the prediction and decision-making of AI algorithms are more correct. However, accessing to a large amount of high-quality training data may not be available in some systems due to privacy and resources constraints\cite{Jordan15}. In addition, the trustworthiness of data also plays an important role. In the exploration of data resources, AI approaches need better data sources for training models to solve the problems more effectively. However, high-accurate and privacy-aware data/intelligence sharing is difficult via the current Internet of information. 

Moreover, although recent advances of AI have made great progress in different applications, most AI algorithms are targeted to learn one specific function from one single data source. By contrast, humans learn from many different types of intelligence and skills. Therefore, lifelong or never-ending AI learners are desirable in the future.

Another aspect of the analogy to human learning is \emph{collective learning}, which is not available in the current AI systems. In understanding the history of cosmos, earth, life and humanity from the Big Bang through to the present, it is collective learning that explains why humans are different from all other animals. With collective learning, humans can learn from data, preserve intelligence, share it with one another, and pass it to the next generation. Collective learning enables humans to adapt to new environments by allowing them to share ideas about how to cope with their surroundings \cite{BigHistory}. 

\subsection{Blockchain Technology}

Recent advances of blockchain technology can help address the challenges described above. With the tremendous development of crypto-currencies, the underlying blockchain has attracted great attentions from both industry and academia \cite{Yu19}. Similar to TCP/IP, which laid the groundwork for the development of the Internet of information, blockchain has great potential to create new foundations for our socio-economic systems by efficiently establishing trust among people and machines, reducing cost, and increasing utilization of resources \cite{IL17}. A blockchain is a continuously growing list of records, called blocks, linked and secured using cryptography. Essentially, it is a consensus of replicated, shared and synchronized data geographically spread across a network of multiple nodes. There is no central administrator or centralized data storage. Using a consensus algorithm, any changes to the ledger are reflected in the copies. The security and accuracy of the ledger are maintained cryptographically according to rules agreed by the network.

\subsection{The Intelligence-Net Enabled by Blockchain}

\begin{figure}[tp]
  \centering
	\vspace{-6 cm}
  \includegraphics[width=1 \textwidth]{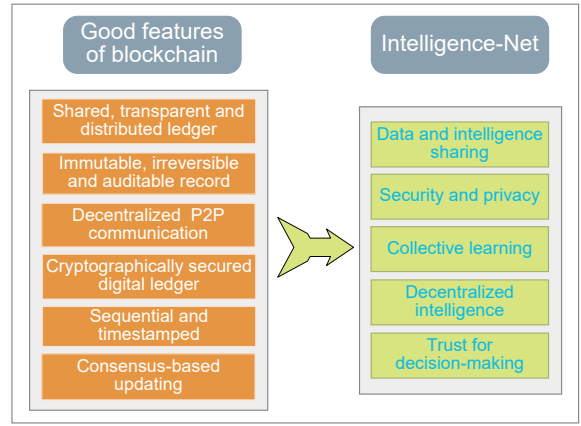}
  \caption{Good features of blockchain that can enable the Intelligence-Net.}
  \label{fig:BlockchainBenefits}
\end{figure}

In this subsection, we show that the good features of blockchain that can enable the Intelligence-Net, including data and intelligence sharing, security and privacy, decentralized intelligence, collective learning, and trust issues. The summary is listed in Figure \ref{fig:BlockchainBenefits}.

\subsubsection{Blockchain can Benefit Data and Intelligence Sharing}

In the current Internet of information and AI systems, the inefficient management of data and intelligence sharing is a key bottleneck of the development of the Intelligence-Net. To deal with these issues, blockchain technology can be used to encourage distributed parties to share data and intelligence with incentive mechanisms embedded in the blockchain. The trustworthiness of data and intelligence plays an important role in the Intelligence-Net. Currently, considering the trust and privacy issues, most users are not willing to share their data and intelligence with others. Taking advantages of the blockchain, it becomes possible to solve this issue. Specifically, every transaction on the blockchain is checked, verified and stored based on the one-way cryptographic hash functions in the distributed network. These ever executed transactions are irreversible and non-repudiable after the verification. Due to these good features of the blockchain, it can enable provenance on data and intelligence, and significantly improve the trustworthiness of the data and intelligence.

\subsubsection{Blockchain can Benefit the Intelligence-Net for Security and Privacy}
Both security and privacy preservation are key factors for using the Intelligence-Net. Currently, most AI systems adopt a centralized architecture, which is prone to hacking and single point of failure. Moreover, since data and intelligence usually involve personal information, the data breaches may lead to the privacy concerns of personal data. How to protect data and intelligence against hackers is an important issue. Thanks to the cryptography embedded in the blockchain, the data stored in the blockchain is highly secure. Blockchain is ideal for storing the highly sensitive, personal data. In addition, punishment schemes embedded in the blockchain can be used to deal with the cyber attack issues. Blockchain can also be used with attribute-based encryption (ABE) technology to enable data and intelligence owners to distribute secret key for users and encrypt shared data in the access control management.

\subsubsection{Blockchain can Benefit the Intelligence-Net for Decentralized Intelligence and Collective Learning}
A centralized AI system needs to have a global vision of the problem and all the necessary knowledge and resources for collecting large datasets. However, the centralized AI system is not adequate for the evolving societal and economic complexity, and can have serious limitations, such as scalability issues in IoT and edge computing applications. Decentralized nature of blockchain can address these issues. By applying blockchain, an individual intelligent entity can learn, train, derive decision-making on local devices in a decentralized and distributed manner. More importantly, intelligence can be preserved and shared with others to enable never-ending and collective learning. In particular, smart contracts in blockchain can provide new opportunities to model the interactions between different AI entities.

\subsubsection{Blockchain can Benefit the Intelligence-Net for Trusting Decision-Making}
As AI systems become smarter and smarter to handle more important tasks, such as autonomous driving, it will become difficult for humans to understand how these AI systems come to specific conclusions and decisions. The lack of trust would severely limit the applications of AI systems. Therefore, the trust mechanisms and audit processes should be well designed in the Intelligence-Net. Blockchain's decentralized, transparent and cryptographic features enable people and machines to trust each other and build a new trust paradigm. With blockchain, the decision-making process can be reviewed and audited at any time by the authorized node. In this way, it provides a feasible solution to audit the process and significantly improves the trustworthiness of the Intelligence-Net.

\section{Open Issues and Research Challenges}
While the research on the Intelligence-Net is still emerging, many open issues and challenges need to be addressed carefully for future efforts. In this section, we discuss some of the open issues and research challenges.

\subsection{Modeling of Intelligence}
In the era of the Internet of information, modeling of information plays a fundamental role. Particularly, Shannon's information theory using ``entropy" to quantify the amount of information has been crucial to the success of the Internet of information. Similarly, modeling of intelligence will be crucial to the success of the Intelligence-Net. Turing test is a popular approach to test a machine's ability to exhibit intelligent behavior equivalent to, or indistinguishable from, that of a human. However, modeling of intelligence is still an open issue. Can we quantify the amount of intelligence using a measure similar to entropy?

\subsection{Architecture and Protocol Designs}
We have gone through layering, cross-layer, and cross-system design paradigms for the Internet of information. In designing the Intelligence-Net, should we go through a similar procedure, or can we consider cross-system design at the first place? Moreover, the Internet of information has the successful hourglass architecture centering on the universal network layer (i.e., IP), which implements the basic functionality necessary for global inter-connectivity. By allowing both lower and upper layer technologies to innovate independently, this ``thin waist" principle has successfully enabled the explosive growth of the Internet of information \cite{ZABJCCPWZ14}. Similarly, we envision a ``thin-waist" architecture of the Intelligence-Net, which needs further research.

\subsection{Scalability of Blockchain and AI Systems}
Currently, most blockchain and AI systems are specialized systems, which are designed for specialized applications. It is very difficult to inter-operate among these systems. In addition, since blockchain was originally designed mainly for crypto-currencies, a number of non-trivial issues in the current blockchain systems prevent it from being used as a generic platform for different services and applications across the globe. Particularly, different services and applications in the Intelligence-Net will have widely varying QoS requirements, which should be fulfilled by advanced QoS provisioning schemes in the future.

\section{Conclusion} \label{sec:conclusion}

In this paper, we reviewed the design approaches in the Internet of information: layering, cross-layer, and cross-system designs. The challenges in the Internet of information motivated us to take a look at humans' cooperation in a larger time scale. Then, we reviewed three major technologies that facilitate humans' cooperation: ``grid of transportation", ``grid of energy", and ``the Internet of information". We observed that each of these three major innovation enables us to move ``something" to reduce the disparity of ``something", by which we can share ``something" to facilitate our cooperation. Based on this observation, we envisioned that the next cooperation paradigm could be ``the Internet of intelligence (Intelligence-Net)", where intelligence can be easily obtained like energy and information. Then, we present blockchain technology and its good features, which can enable the Intelligence-Net. Finally, open issues and research challenges about the Intelligence-Net were outlined and discussed.

\ifCLASSOPTIONcaptionsoff
  \newpage
\fi

%\section*{Acknowledgment}
%We thank the reviewers for their detailed reviews and constructive comments, which have helped to greatly improve the quality of this paper.

\balance

\bibliographystyle{IEEEtran}
\bibliography{XieReferences,D:/CA/Papers/Ref}

\begin{thebibliography}{10}
\providecommand{\url}[1]{#1}
\csname url@samestyle\endcsname
\providecommand{\newblock}{\relax}
\providecommand{\bibinfo}[2]{#2}
\providecommand{\BIBentrySTDinterwordspacing}{\spaceskip=0pt\relax}
\providecommand{\BIBentryALTinterwordstretchfactor}{4}
\providecommand{\BIBentryALTinterwordspacing}{\spaceskip=\fontdimen2\font plus
\BIBentryALTinterwordstretchfactor\fontdimen3\font minus
  \fontdimen4\font\relax}
\providecommand{\BIBforeignlanguage}[2]{{%
\expandafter\ifx\csname l@#1\endcsname\relax
\typeout{** WARNING: IEEEtran.bst: No hyphenation pattern has been}%
\typeout{** loaded for the language `#1'. Using the pattern for}%
\typeout{** the default language instead.}%
\else
\language=\csname l@#1\endcsname
\fi
#2}}
\providecommand{\BIBdecl}{\relax}
\BIBdecl

\bibitem{CLC07}
M.~{Chiang}, S.~H. {Low}, A.~R. {Calderbank}, and J.~C. {Doyle}, ``Layering as
  optimization decomposition: A mathematical theory of network architectures,''
  \emph{Proceedings of the IEEE}, vol.~95, no.~1, pp. 255--312, Jan. 2007.

\bibitem{SM05}
V.~Srivastava and M.~Motani, ``Cross-layer design: a survey and the road
  ahead,'' \emph{IEEE Comm. Mag.}, vol.~43, no.~12, pp. 112--119, Dec. 2005.

\bibitem{AEM16}
I.~{Al-Anbagi}, M.~{Erol-Kantarci}, and H.~T. {Mouftah}, ``A survey on
  cross-layer quality-of-service approaches in {WSNs} for delay and
  reliability-aware applications,'' \emph{IEEE Comm. Surveys Tutorials},
  vol.~18, no.~1, pp. 525--552, Firstquarter 2016.

\bibitem{FGA08}
F.~{Foukalas}, V.~{Gazis}, and N.~{Alonistioti}, ``Cross-layer design proposals
  for wireless mobile networks: A survey and taxonomy,'' \emph{IEEE Comm.
  Surveys Tutorials}, vol.~10, no.~1, pp. 70--85, First Quarter 2008.

\bibitem{CYH16}
Q.~Chen, F.~R. Yu, T.~Huang, R.~Xie, J.~Liu, and Y.~Liu, ``An integrated
  framework for software defined networking, caching and computing,''
  \emph{IEEE Network}, vol.~31, no.~3, pp. 46--55, May 2017.

\bibitem{YHL18}
T.~H. F.~Richard~Yu and Y.~Liu, \emph{Integrated Networking, Caching and
  Computing}.\hskip 1em plus 0.5em minus 0.4em\relax CRC Press, 2018.

\bibitem{WHY18}
C.~Wang, Y.~He, F.~R. Yu, Q.~Chen, and L.~Tang, ``Integration of networking,
  caching and computing in wireless systems: A survey, some research issues and
  challenges,'' \emph{IEEE Comm. Surveys Tutorials}, vol.~20, no.~1, pp. 7--38,
  Firstquarter 2018.

\bibitem{YH18}
F.~R. Yu and Y.~He, \emph{Deep Reinforcement Learning for Wireless
  Systems}.\hskip 1em plus 0.5em minus 0.4em\relax Springer, 2018.

\bibitem{Jordan15}
M.~I. Jordan and T.~M. Mitchell, ``Machine learning: Trends, perspectives, and
  prospects,'' \emph{Science}, vol. 349, no. 6245, pp. 255--260, 2015.

\bibitem{FXD14}
B.~{Fu}, Y.~{Xiao}, H.~J. {Deng}, and H.~{Zeng}, ``A survey of cross-layer
  designs in wireless networks,'' \emph{IEEE Comm. Surveys Tutorials}, vol.~16,
  no.~1, pp. 110--126, Firstquarter 2014.

\bibitem{SB18}
R.~S. Sutton and A.~G. Barto, \emph{Reinforcement Learning: An Introduction,
  Second Edition}.\hskip 1em plus 0.5em minus 0.4em\relax MIT Press, 2018.

\bibitem{GBC16}
I.~Goodfellow, Y.~Bengio, and A.~Courville, \emph{Deep Learning}.\hskip 1em
  plus 0.5em minus 0.4em\relax MIT Press, 2016,
  \url{http://www.deeplearningbook.org}.

\bibitem{Harari14}
Y.~N. Harari, \emph{Sapiens: A Brief History of Humankind}.\hskip 1em plus
  0.5em minus 0.4em\relax Harper, 2014.

\bibitem{MNV09}
K.~Maruyama, F.~Nori, and V.~Vedral, ``Colloquium: The physics of maxwell's
  demon and information,'' \emph{Rev. Mod. Phys.}, vol.~81, pp. 1--23, Jan.
  2009.

\bibitem{BAP12}
A.~Bérut, A.~Arakelyan, A.~Petrosyan, S.~Ciliberto, R.~Dillenschneider, and
  E.~Lutz, ``Experimental verification of landauer’s principle linking
  information and thermodynamics,'' \emph{Nature}, vol. 483, no. 7388, pp.
  187--190, 2012.

\bibitem{BigHistory}
``Big history project,'' website: www.bighistoryproject.com.

\bibitem{Yu19}
F.~R. Yu, ``A service-oriented blockchain system with virtualization,''
  \emph{Trans. Blockchain Technology and Applications}, vol.~1, no.~1, pp.
  1--10, Firstquarter 2019.

\bibitem{IL17}
M.~Iansiti and K.~R. Lakhani, ``The truth about blockchain,'' \emph{Harvard
  Business Review}, Jan. 2017.

\bibitem{ZABJCCPWZ14}
{L.~Zhang}, {A.~Afanasyev}, {F.~Burke}, {V.~Jacobson}, {K.C.~Claffy},
  {P.~Crowley}, {C.~Papadopoulos}, {L.~Wang}, and {B.~Zhang}, ``Named data
  networking,'' \emph{ACM SIGCOMM Comput. Commun. Rev.}, vol.~44, no.~3, pp.
  66--73, Jul.~2014.

\end{thebibliography}

\end{document}